\documentclass[aps,twocolumn,showpacs]{revtex4}
\usepackage{graphicx}%
\usepackage{amsmath,amssymb}

\begin{document}
\title{Nonlinear structures and thermodynamic instabilities in a
one-dimensional lattice system
}
\author{Nikos Theodorakopoulos$^{1,2}$, Michel Peyrard$^{3}$ and Robert S. MacKay$^{4}$} 
\affiliation{
$^{1}$Theoretical and Physical Chemistry Institute, National Hellenic Research Foundation,
Vasileos Constantinou 48, 116 35 Athens, Greece\\
$^{2}$Fachbereich Physik, Universit\"at Konstanz, 78457 Konstanz, Germany\\
$^{3}$Laboratoire de Physique, UMR-CNRS 5672, ENS Lyon, 
46 All\'{e}e d'Italie, 69007 Lyon, France\\
$^{4}$Mathematics Institute, University of Warwick, Coventry CV4 7AL, U.K.
}
\date{\today}
\begin{abstract}
The equilibrium states of the discrete Peyrard-Bishop Hamiltonian with one end fixed are computed exactly
from the two-dimensional nonlinear Morse map. These exact nonlinear
structures are interpreted as domain walls (DW),
interpolating between bound and unbound segments of the chain. 
The free energy of the DWs is calculated to leading order beyond the Gaussian approximation. 
Thermodynamic instabilities
(e.g. DNA unzipping and/or thermal denaturation) can be understood in terms of DW
formation. 
\end{abstract}
\maketitle

Computational results \cite{numvar} 
have 
suggested, 
at varying levels of numerical rigor, 
that a class of one-dimensional lattice models proposed within a variety of physical contexts
(e.g. interfacial wetting \cite{KroLip} or DNA denaturation \cite{PB}) 
may exhibit thermodynamic instabilities \cite{B+}
which share many of the properties of ordinary phase transitions.  

Owing to the lack of analytical results,
it is important to explore alternative strategies which reveal 
exact features of these instabilities.
One such strategy, which will be pursued in this work,
is to exploit some of the tools of nonlinear dynamics in order to study the properties of the underlying 
exact equilibrium structures. 


The general class of models under consideration is described by 
a Hamiltonian with 
a configurational part 
\begin{equation}
\Phi  = \sum_{n=0}^{N}\Biggl[ 
  \frac{1}{ 2R} (y_{n+1}-y_{n})^{2} + V(y_n)   \Biggr]
\label{eq:PBHam}
\end{equation}
where $y_n$ is the transverse displacement of the $n$th site, $R$ is a dimensionless
coupling constant  and $V(y)$ is any 
potential with a  repulsive core, a stable minimum and a flat top. 
We will deal with the case of  a Morse potential \cite{PB}, 
$V(y)= (1 - e^{-y})^{2} $. All quantities referred to in this paper are dimensionless.

We will demonstrate the existence of exact nonlinear structures which
correspond to domain walls (DW)s, \lq\lq interpolations\rq\rq\- from the bound to the unbound phase. It
will further be shown that the detailed stability properties of the DWs essentially 
determine the system's behavior with respect to both mechanical and thermal 
instabilities (unzipping and thermal melting, respectively, in the DNA context).  
Finally, we will use thermodynamic perturbation theory to account for the
minor discrepancies which occur between the predictions of Gaussian 
DW-based theory and standard numerical results
based on the transfer integral (TI) method. 
This work describes the case $R\gg1$ (extreme discretization);  
DWs in the case $R \ll 1$ (continuum limit) 
have been treated in Ref. \cite{DTP}. 
Unless otherwise stated, the
value $R=10.1$ will be used in numerical applications.\par
\begin{figure}[h]
\vskip -.5truecm
\includegraphics[width=65mm]{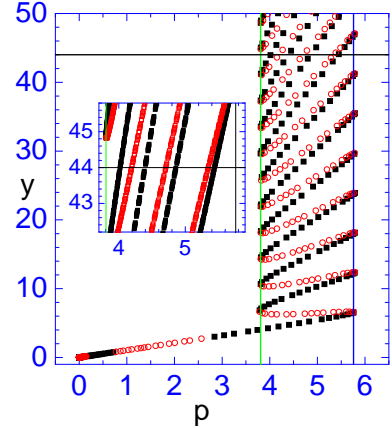}
\vskip -.5truecm
\caption{ 
{\small 
The unstable manifold of the FP of the map (\ref{eq:map})
for $R=10.1$ and $N=28$. 
Black squares belong to stable equilibria, red open circles belong to unstable equilibria.
The horizontal line at $y=44$ demostrates the multivaluedness of
the manifold as a function of $y$ (4 stable and 3 unstable equilibria with
that value of $y$; details in the inset). 
The vertical lines are drawn at $p_{min}$ and $p_{max}$, the  minimal and
maximal asymptotic slopes of DWs.
}
}
\label{fig:manif}
\end{figure}
The equilibria $\{y_n^{(\alpha)}\}$ of (\ref{eq:PBHam}),
subject to fixed-end boundary conditions $y_0, y_{N+1} = const$,
are the solutions of the two-point boundary value problem for the 
second order recurrence relation $y_{n+1}-2y_n+y_{n-1} = RV'(y_n)$. 
By introducing $p_n=y_n-y_{n-1}$, this is equivalent to orbits
of the two-dimensional map 
\begin{eqnarray}
\nonumber
p^{(\alpha)}_{n+1}&=&p^{(\alpha)}_{n}+RV'(y_{ n}^{(\alpha)})\\
y_{n+1}^{(\alpha)}&=&y_{n}^{(\alpha)}+p_{n+1}^{(\alpha)}  \quad,
\label{eq:map} 
\end{eqnarray}
where $ n=1,\cdots, N $, 
$y_1=p_1+y_0$, and $p_1$ is unspecified, i.e. it can take 
any value which leads to the given $y_{N+1}$ after $N$ iterations.
The map (\ref{eq:map})
has a single hyperbolic fixed point (FP) at $(p^{(0)}=y^{(0)}=0)$.
Equilibria can be classified as stable or unstable, according to 
whether all eigenvalues of the Hessian
\begin{equation}
A_{mn}^{(\alpha)}  = \left.  \frac{\partial^2 \Phi }{ \partial y_n \partial y_m }
\right|_{\{y_i = y_{i}^{(\alpha)} \}}
\label{eq:Hessian}
\end{equation}
are nonnegative or not.

The tangent map at the FP is given by 
\begin{equation}
\left(
\begin{array}{c}
\delta p \\
\delta y
\end{array}
\right)_{ n+1}
=
\left(
\begin{array}{cc}
1   & 2R  \\
 1  &   1+2R 
\end{array}
\right)  
\left(
\begin{array}{c}
\delta p \\
\delta y
\end{array}
\right)_{ n}
   \>;
\label{eq:Tmap}
\end{equation}
its eigenvalues are $ \lambda _{ \pm }=1+R\pm \sqrt{R^{2}+2R}$
and its corresponding (unnormalized) eigenvectors 
${\vec \kappa }^{( \pm)}=\left( 2R, R \pm \sqrt{R^{2}+2R } \right)$,
where the upper (lower) sign corresponds to the unstable (stable) manifold.
\begin{figure}
\vskip -.5truecm
\includegraphics[width=65mm]{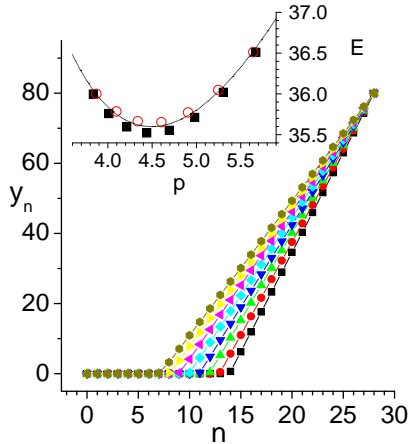}
\vskip -.5truecm
\caption{ 
{\small 
The 8 stable equilibria corresponding to $N=28$,  $y_0=0$, $ y_{N+1}=L=80$.
Not shown are 7 unstable equilibria enmeshed between the stable ones. Inset:
total energies for both stable (black squares)
and unstable (red open circles) equilibria.
The continuous curve corresponds to a theoretical estimate which does not
distinguish between stable and unstable equilibria (cf. text). Note that the 
typical energy difference - which does not shrink  with increasing $L$ - 
between neighboring extrema 
is about $0.15$.
}
}
\label{fig:DW}
\end{figure}
The structure of the unstable manifold of the FP is shown in Fig. \ref{fig:manif}. 
Its multivaluedness
as a function of $y$ turns out to
have important consequences when the fixed-ends boundary conditions appropriate
to the problem are introduced. 
These are of two types: (a) $y_0= y_{N+1}=0$, and (b)
$y_0 = 0, y_{N+1}=L$. The only equilibrium compatible with (a) is the FP.
In order to construct the latter,
we generated a large number of sequences using the initial conditions
$y_1=p_1= \kappa ^{+}_2/\kappa ^{+}_1 \> \exp(-s) $ with $s$ uniformly 
spaced in an interval $(s_1,s_2)$ and initially accepted
a sequence if $|y_{N+1}-L|/L<2\cdot 10^{-5}$; a further scan of accepted sequences
was made to eliminate sequences which are \lq\lq immediate neighbors\rq\rq\- in the manifold.
The result  for
$L=80$ is a total of   8 stable  and  7 unstable such sequences. 
They have the form of DWs consisting of a segment bound in the well 
of the Morse potential connected to a free segment on its plateau.
The stable ones are  
shown in Fig. \ref{fig:DW}; at the left they more or less coincide with the fixed point $y^{(0)}=0$;
to the right, they emerge as a bundle with linearly 
growing displacements and a range of slopes $p_{min}\le p_{\alpha } \le p_{max}$.
The corresponding energies, according to
(\ref{eq:PBHam}) are shown in the inset as functions of the final slope $p$.
Note that the unstable equilibria have energies slightly above the neighboring
stable ones. The overall dependence of the energy on the slope $p$ can be understood by the following
simple argument: each unbound site contributes an amount $1+p^2/(2R)$ to the energy; for $L\gg 1$, 
the number of unbound sites 
is approximately equal to $L/p$; 
hence $E(p,L) = (1/p + p/2R) L$; this is the dotted curve shown 
in the inset of Fig.  \ref{fig:DW}. It follows that the minimum energy 
occurs at 
$p=p^{*}=(2R)^{1/2}$ and has value $E^{*}(L) = 2 L / p^{*}$.


A few remarks are in order here: (i) The number of equilibria grows linearly with $L$;
more precisely, for sufficiently large $L$, the number of local minima
is $[ L/p_{min}] - [ L/p_{max} ] + 1$; there is one 
fewer unstable equilibrium.
(ii) There are no zero eigenvalues 
for any equilibria, except for the discrete values of
$L$ at which a new pair is 
created. (iii) 
The unstable equilibria have only one negative eigenvalue, so we call them saddles;
the rest of their spectrum is essentially 
identical with that of neighboring local minima. 
(iv) For large $R$, the 
limits $p_{min}, p_{max}$ are asymptotically $p_{min} \sim \ln (2R) +1, \> p_{max} \sim 
R/2 + \ln2$, which we derived by approximate solution of the equations for 
the unstable manifold. 
(v)  The energy spacing (Peierls barrier) between the absolute minimum 
and the saddle with the lowest energy
(cf. inset in Fig. \ref{fig:DW}) is about 0.15.  
(vi) For any given equilibrium (DW) 
with a transverse displacement $L\gg 1$ and final slope $p$, the quantity
 $ dE / dL  =  V^{'}(L) + (L-y_{N})/R \approx p/R $ represents
the force acting on the end particle. In the case of the absolute minimum, $p=p^{*}$,
this is equal to $f_{0}=(2/R)^{1/2}$; in the context of DNA denaturation
models, it corresponds to the force required to bring about mechanical \lq\lq unzipping" at 
zero temperature.  

What happens at nonzero temperatures? In order to examine the relevance of DWs 
for thermodynamics, it is necessary to look at the partition functions 
$Z_N(0)$ and $Z_N(L)$ which correspond to the two types of boundary conditions above. 
The quantity of interest is the difference in the free energies
\begin{equation}
\Delta G = -T  \lim _{N \to \infty }\ln \left\{ \frac{Z_N(L)}{Z_N(0) }\right\} \quad ;
\label{eq:DG}
\end{equation} 
in particular, the derivative $ \left( \partial \Delta G / \partial L \right)_{T} $ 
is equal to the force $f(L,T)$ required to maintain the right end of 
the chain at a given tranverse displacement $L$.
For small displacements around any equilibrium,
$y_n 
= y_n^{(\alpha)} + \psi _n$, $\Phi $ can be 
expanded as
\begin{eqnarray}
\nonumber
\Phi^{(\alpha )}(\{\psi  \}) & \approx  & E^{(\alpha )} +
 \frac{1}{2}\sum_{m,n}^{} A_{mn}^{(\alpha )}\psi_m\psi_n\\
&  & +  \sum_{m}^{}v_{m,\alpha} ^{(3)} \psi_{m}^{3}
+  \sum_{m}^{}v_{m,\alpha} ^{(4)} \psi_{m}^{4} \>,
\label{eq:Hquartic}
\end{eqnarray}
where $v_{m,\alpha} ^{(j)} = V^{(j)}(y_{m}^{(\alpha )})/j!$\-,
and terms of higher than fourth order have been dropped.
The partition function 
\begin{equation}
Z_{N}^{(\alpha)} = 
\int_{ -\infty }^{+\infty  }
\prod_{j=1}^{N} d\psi _{j}  \>\>
e^{- \Phi^{(\alpha )}(\{\psi  \})/T }
\label{eq:Za}
\end{equation}
can be calculated in the Gaussian approximation by keeping
terms up to second order in (\ref{eq:Hquartic}). The result is 
\begin{equation}
Z_{N}^{(\alpha)} \approx 
   e^{- E^{(\alpha)}/T}
 \prod_{\nu } \left\{ 2\pi T /  \Lambda _{\nu }^{(\alpha) }    \right \} ^{1/2} 
\quad ,
\label{eq:Zalpha}
\end{equation}
where $\{\Lambda _{\nu }^{(\alpha)} \}$, $\nu =1,2, ..., N$ are the eigenvalues of the Hessian matrix 
$A^{(\alpha)}$.

Eq. (\ref{eq:Zalpha}) will be applied twice. First, to evaluate $Z_N(0)$; in this case, the relevant 
equilibrium is the FP. Second, to evaluate $Z_N(L)$, in which case the
relevant equilibrium is the DW with the minimal energy and the final slope $p^{*}$ (cf. above). 
Note that at temperatures of order 1 or higher (relevant for the occurrence 
of macroscopic instabilities), the roughness of the energy 
landscape (Fig. \ref{fig:DW}), 
typified by the height of the Peierls barrier (cf. comment (v) above),
essentially disappears. 

The ratio of partition functions in (\ref{eq:DG}) can be approximated as
\begin{equation}
\frac{Z_N(L)}{Z_N(0) }\approx  
 e^{- E^{(*)}/T}
\prod_{\nu } \left\{ \Lambda _{\nu }^{(0) } /  \Lambda _{\nu }^{(*) }  \right \} ^{1/2} 
\quad.
\label{eq:Zratio}
\end{equation}
\begin{figure}
\vskip -.5truecm
\includegraphics[width=65mm]{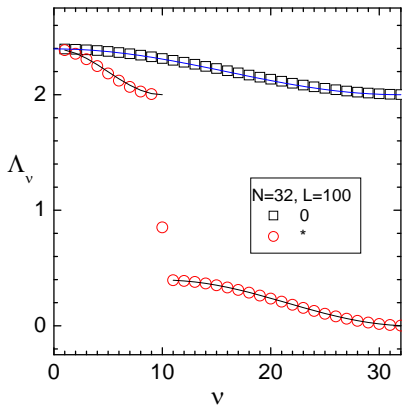}
\vskip -.5truecm
\caption{ 
{\small 
Eigenvalue spectra of the hessians $A^{(0)}$ (open squares) 
and $A^{(*)}$ (open circles)
for $N=32$ and $L=100$. The DW's spectrum consists
of bands of optical and acoustic phonons, localized respectively
in the bound and unbound portions of the chain,
and a single local mode in the gap; both bands
are well described (to order ${\cal O}(1/L)$) by the corresponding
free phonon dispersion curves (dotted).
}
}
\label{fig:spectra}
\end{figure}
The eigenvalue spectra $\{ \Lambda _{\nu }^{(*) }, \Lambda _{\nu }^{(0) }
  \}$ are shown in Fig. \ref{fig:spectra}. As has already been pointed out in the context of the 
continuum-limit\cite{DTP}, the DW acts as an interface between bound and unbound phases;
the unbound phase supports only acoustic phonons and the bound phase only optical ones.
The quantity
\begin{equation}
\Omega =  \frac{1}{2} \sum_{\nu=1 }^{N} 
\ln \left\{ \Lambda _{\nu }^{(0) } /  \Lambda _{\nu }^{(*) }    \right \} 
= \frac{L}{p^{*}} \sigma   +  {\cal O} (1) \>,
\label{eq:entropy}
\end{equation}
where $\sigma = \ln\left(\sqrt{R/2}  + \sqrt{1+ R/2}\right)$,
is found to be, for $L/p^{*}\gg 1$ and $N-L/p^{*}\gg 1$, independent of $N$
and proportional to the number $L/p^{*}$
of unbound sites; the reason is that, as can be seen from Fig. \ref{fig:spectra}, the optical phonons in the bound
region of the DW cover exactly the frequency range of the optical phonons around the fixed point, and therefore cancel out
in taking the ratio; this leaves only a dependence on the ratio between acoustic and optical
frequencies for a number of modes of order $L/p^{*}$. 
The proportionality constant $\sigma (R) $ has been calculated in Ref. \cite{DTP} using the argument
sketched above, without reference to the exact eigenvalue spectrum. 
Irrelevant corrections of order unity originate from the transition region of
the DW and/or the single localized oscillation mode. Inserting (\ref{eq:entropy})
and (\ref{eq:Zratio}) in (\ref{eq:DG}), we obtain, in the Gaussian approximation,
\begin{equation}
\Delta G  \approx E^{*}(L) - T\Omega  =  \left(  2  -  T \sigma  \right)  \frac{L}{p^{*} } \quad.
\label{eq:DGGauss}
\end{equation}
It follows that the force now required to maintain the chain at (any) transverse displacement
$L\gg p^{*}$,
\begin{equation}
f(T) \approx   \frac{1}{p^{*} }  \left( 2  -   T \sigma \right)  \quad,
\label{eq:TdepFGauss}
\end{equation}
is reduced as a result of thermal motion. At $T= 2/\sigma$ this generalized
unzipping force vanishes;
spontaneous formation of the DW becomes possible. 
In the context of DNA denaturation theory, this amounts to spontaneous melting (thermal denaturation).
That transition has been extensively studied, assuming periodic boundary conditions,
both in the continuum and the discrete regime  by means of
the TI method. 
The critical temperature $2/\sigma $ obtained in the Gaussian approximation
is in good agreement with the TI results for values of  $R \lesssim 4$;
systematic deviations occur at higher $R$ however\cite{DTP}.  

\begin{figure}
\vskip -.5truecm
\includegraphics[width=65mm]{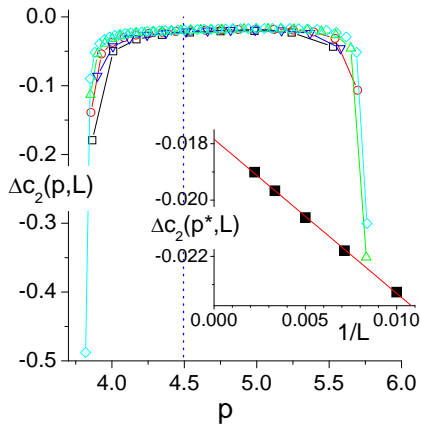}
\vskip -.5truecm
\caption{ 
{\small  
Lowest order anharmonic correction to the free energy difference $\Delta G$:
The coefficient $\Delta c_{2}(p,L)$ for $L=100, 140, 200, 300, 450$ vs. $p$;
for each $L$ the difference  $(C_{2}(p)-C_{2}^{0})/L$ is numerically evaluated at
all stable minima; the interpolates taken at $p=p^{*}$ are plotted against
$1/L$ (inset); this procedure extrapolates to a value $\Delta c_{2}^{*}=-0.01784(5)$.
}
}
\label{fig:Dc2}
\end{figure}
\begin{figure}
\vskip -.5truecm
\includegraphics[width=65mm]{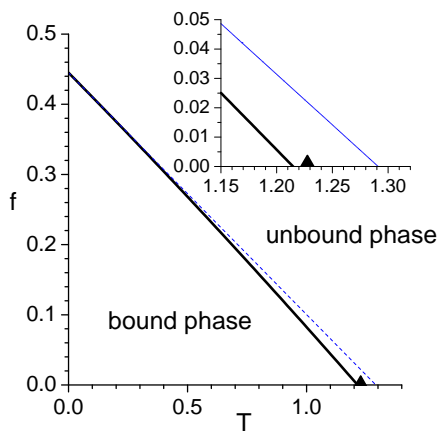}
\vskip -.5truecm
\caption{ 
{\small 
Phase diagram in the $f-T$ plane. The dotted
line is obtained on the basis of the Gaussian approximation (predicted $T_c=1.291$).
The solid line is Eq. \ref{eq:funzipanh}
with $\Delta c_{2}^{*}=-0.01784$ (cf. Fig. \ref{fig:Dc2}). Inset: detail of 
critical region, showing a predicted $T_c=1.215$; the triangle is the result of a finite-size scaling
numerical TI calculation $T_c=1.227$ [\onlinecite{numvar}c].
}
}
\label{fig:PhaseDiagram}
\end{figure}
It is possible to improve the calculation of $\Delta G$ by going beyond the Gaussian approximation.
Standard thermodynamic perturbation theory\cite{Amit} can be applied  to  evaluate 
the contribution of cubic and quartic terms in (\ref{eq:Hquartic}) 
to the partition function (\ref{eq:Za}),
- again, both for the FP and the DW - and to derive a low-temperature expansion
for the resulting free energy.
Each power of the displacement field in (\ref{eq:Hquartic}) 
contributes a factor $T^{1/2}$.
The  irreducible graphs which contribute to $\ln Z_N^{(\alpha )}$ to leading order in
the temperature are: %
\begin{eqnarray}
\bigcirc \hspace{-2pt} { \blacksquare} \hspace{-2pt} \bigcirc & =& 
- 3T \> \sum_{m}^{}v_{m,\alpha} ^{(4)} \{g_{mm}^{\alpha} \}^{2} \quad,{\rm and}\\
\blacktriangleright  \hspace{-5pt}
\frac{\frown}{\smile }
\hspace{-4pt}   \blacktriangleleft \>
&= &\>
 6T \cdot \frac{1}{2 }   \sum_{m}^{}
v_{m,\alpha} ^{(3)} v_{n,\alpha} ^{(3)} 
\{g_{mn}^{\alpha } \}^{3}\quad ,
\label{eq:term34}
\end{eqnarray}
where $g_{mn}^{(\alpha )} $ is the relevant Green's function (inverse of the 
Hessian evaluated at the equilibrium $\alpha $).

Let $-TC_2^{\alpha }$ denote the sum of the two contributions above.
The correction to the free energy $\Delta G$ will be 
obtained as a difference between values at the two relevant equilibria
(FP and DW with minimal energy); the quantity $C_{2}^{(*)} - C_{2}^{(0)}$
is expected to be independent of $N$ and proportional to
the length of the unbound segment (cf. above).  Fig. \ref{fig:Dc2} (inset) displays the
values of $\Delta c_2(p^{*}) = (C_{2}^{(*)} - C_{2}^{(0)})/L$ (interpolated from the 
values at the closest DW's, displayed in the main figure) for a series of $L$-values. 
The extrapolated
value $\Delta c_{2}^{*} = \lim_{L \to \infty }\Delta c_2 (p^{*},L)$
can be used as a correction to eqs.  \ref{eq:DGGauss} and
\ref{eq:TdepFGauss}
above, i.e.
\begin{equation}
f(T) \approx   \frac{1}{p^{*} }  \left(  2  -   T \sigma 
 \right)  + \Delta c_{2}^{*}  \> T^2   \quad.
\label{eq:funzipanh}
\end{equation}
The resulting phase diagram is shown in Fig. \ref{fig:PhaseDiagram}.
The predicted $T_c=1.215$ is in good agreement with the numerical 
TI calculation [\onlinecite{numvar}c]
$T_{c}^{TI}=1.227$. \par

We conclude with two remarks. The first is a caution about the inherent limitations of
the low-order perturbational approach used. At lower values of $R$, which tend to increase
$T_c$ to values substantially higher than 1, it cannot be expected to work (although the Gaussian
approximation produces very good estimates of $T_c$). At much higher values of
$R$, the perturbation terms in the Hamiltonian 
become too large compared with the Gaussian approximation; 
again, one would presumably have to include higher order terms to obtain reasonable results.\par

The second remark concerns a further prospect: including nonlinear stacking terms in
the Hamiltonian \cite{TDP} has been known to produce real or apparent first order transitions. It would be interesting
to explore how this behavior would be reflected in the DW approach.\par

Part of this work was performed while one of us (N.T.) was visiting Lyon and Warwick; he wishes to thank
both departments for hospitality. Financial support from 
EU contract HPRN-CT-1999-00163 (LOCNET network) is acknowledged.
\par

\end{document}